\documentclass[aps,prd,twocolumn,nofootinbib,superscriptaddress]{revtex4-2}

\usepackage[utf8]{inputenc}
\usepackage{amsmath,amsfonts,amssymb,slashed}  
\usepackage{booktabs}
\usepackage{enumitem}
\usepackage{graphicx} 
\usepackage[colorlinks=true,urlcolor=black,citecolor=blue,linkcolor=black] {hyperref}
\usepackage{enumitem}
\setlist[itemize]{leftmargin=*}

\newcommand{\eg}{{\it e.g.}}

\newcommand{\be}{\begin{equation}}
\newcommand{\ee}{\end{equation}}
\newcommand{\br}{\begin{eqnarray}}
\newcommand{\bea}{\begin{eqnarray}}
\newcommand{\eea}{\end{eqnarray}}
\newcommand{\er}{\end{eqnarray}}
\newcommand{\ba}{\begin{array}}
\newcommand{\ea}{\end{array}}
\newcommand{\bi}{\begin{itemize}}
\newcommand{\ei}{\end{itemize}}
\newcommand{\bn}{\begin{enumerate}}
\newcommand{\en}{\end{enumerate}}
\newcommand{\bc}{\begin{center}}
\newcommand{\ec}{\end{center}}

\newcommand{\SU}{{\rm SU}}

\newcommand{\Tr}{\,{\rm Tr}}

\newcommand{\nicpb}{Laboratory of High Energy and Computational Physics, NICPB, R\"avala pst. 10, 10143 Tallinn, Estonia}
\newcommand{\helsinki}{Department of Physics, and Helsinki Institute of Physics, P.O. Box 64, FI-00014 University of Helsinki, Finland}
\newcommand{\bennett}{Department of Physics, Bennett University, Plot Nos 8-11, TechZone II, Greater Noida 201310, Uttar Pradesh, India}

\newcommand{\beq}{\begin{equation}}
\newcommand{\eeq}{\end{equation}}

\begin{document}

\title{Multi-phase critical Higgs boson at colliders}

\author{Katri Huitu}
\affiliation{\helsinki}

\author{Kristjan Kannike}
\affiliation{\nicpb}

\author{Niko Koivunen}
\affiliation{\nicpb}

\author{Luca Marzola} 
\affiliation{\nicpb}

\author{Subhadeep Mondal}
\affiliation{\bennett}

\author{Martti Raidal}
\affiliation{\nicpb}

\date{\today}

\begin{abstract}
The recently proposed multi-phase criticality principle in Coleman-Weinberg models can provide a new explanation for the hierarchy between the electroweak and new physics scales. When applied to the Standard Model, a Higgs boson as light as the pseudo-Goldstone boson of broken scale invariance occurs. The suppressed mixing between the two light fields still carries information about the large scale of symmetry breaking, albeit up to logarithmic corrections. In this work we probe this scenario with the present LHC data and assess the impact of future lepton and hadron colliders. Our results show that the multi-phase criticality can easily explain the apparent absence of new physics at the energy scales tested in current experiments.
\end{abstract}

\maketitle

\section{Introduction}

Proposed solutions to the hierarchy problem, such as supersymmetry~\cite{Wess:1974tw,Fayet:1977vd}, predict a plethora of new particles
at electroweak scale with couplings that cancel quadratic divergencies to the Higgs boson mass~\cite{Chatrchyan:2012ufa,Aad:2012tfa}.
However, the absence of these particles at energies below 5-7~TeV was indicated already by the LEP precision data~\cite{Barbieri:1999tm,Barbieri:2000gf}, thereby creating the little hierarchy problem. 

This triggered a significant theoretical effort aimed at pushing the solution to the hierarchy problem to some higher scale while keeping the Higgs boson naturally light. Frameworks like ``the little Higgs"~\cite{ArkaniHamed:2001nc,ArkaniHamed:2002qy,Cheng:2003ju} or ``the twin Higgs"~\cite{Chacko:2005pe} were developed, which rely on {\it different} sets of new particles to cancel the quadratically divergent contributions to the Higgs mass at one-loop level only, creating a mass gap for the little hierarchy. Unfortunately, the accumulated experimental results from the LHC have shown that no new particles with couplings of order unity to the Standard Model (SM) exist, pushing the scale of any such a framework above several TeV, where they cannot be considered natural.

It was recently proposed~\cite{Kannike:2021iyh} that the lightness of Higgs boson and the apparent absence of any associated new particle can
simultaneously be explained by multi-phase criticality in dynamical symmetry breaking {\it \`{a} la}  Coleman and Weinberg~\cite{Coleman:1973jx}.
At the critical point, where two different symmetry-breaking phases are smoothly connected, the Higgs boson mass is suppressed by loop factors similarly
to that of the dilaton -- the pseudo-Goldstone boson of broken scale invariance. The Higgs and dilaton masses are independent, suppressed by different $\beta$-functions,
and their mixing is also suppressed.  This effect can be understood as a consequence of a little misalignment that quantum corrections induce between the particular tree-level flat direction indicated by the Gildener-Weinberg method~\cite{Gildener:1976ih} and the actual direction where the minimum is generated by radiative corrections, which lies in a different but smoothly connected phase of the theory.

At the electroweak scale, only the Higgs boson and the weakly mixed dilaton appear, while the new physics inducing the dynamical symmetry breaking decouples. 
The generated hierarchy depends on unknown scalar quartic couplings which, unlike in the case of supersymmetry, can have arbitrary values. The existing experimental information indicates that the magnitude of scalar quartic couplings can be vastly different.  For instance, the Higgs boson quartic $\lambda_H$  is large at low energies, but runs to very small values at high energy scales of the order of $10^{10-12}$~GeV~\cite{Degrassi:2012ry,Buttazzo:2013uya}.\footnote{The possible appearance of two degenerate minima in the SM Higgs potential~\cite{Froggatt:1995rt}, which sometimes is also called criticality, should not be confused with our framework of multi-phase criticality.} At the same time, the inflaton~\cite{Linde:1983gd} must have a self-coupling smaller than $10^{-13}$~\cite{Bezrukov:2007ep} in order to comply with the Cosmic Microwave Background measurements~\cite{Akrami:2018odb}.  
As we will show, the proposed multi-phase criticality framework can easily produce hierarchies relevant for the little hierarchy problem by simply using the range of quartic couplings seemingly allowed by Nature.

The aim of this work is to formulate the effective low energy theory supported by the multi-phase criticality scenario applied to the SM and to work out its phenomenology at lepton and hadron colliders. We show that the low energy observables -- the Higgs boson mass, the dilaton mass and their mixing angle -- can determine the scale $\Lambda$ of new physics, identified with the vacuum expectation value (VEV) of the singlet scalar $s$ which triggers the dynamical symmetry breaking, up to a model dependent logarithmic correction $\ln R.$ This correction measures the deviation of the location of the true minimum from the Gildener-Weinberg approximation and is predicted to be small in any weakly coupled realization of the mechanism. After proposing an effective model of the multi-phase critical Higgs boson, we derive the bounds that present and future collider experiments give on the parameter space of the associated dilaton. We then use these constraints to infer a lower bound on the large scale where the symmetry breaking dynamics takes place, corresponding to the scale where the decoupled new degrees of freedom should appear.

\section{Low-energy effective model for multi-phase critical Higgs}
\label{critical}

In order to study the collider phenomenology of the multi-phase critical Higgs boson we first formulate the minimal effective low energy model. This allows us to study the predictions of the framework and to determine the scale of new physics independently of the high-energy dynamics driving the Coleman-Weinberg symmetry breaking. A light Higgs boson arises at the intersection of two phases, in one of which $\SU(2)_L$ must be spontaneously broken. 

Consider a minimal model with two scalar fields: the Higgs doublet
$H = (0,h/\sqrt{2})$ and a neutral singlet scalar $s$, with the biquadratic potential
\beq 
\begin{split}
V &= \lambda_H |H|^4 +\lambda_{HS} |H|^2 \frac{s^2}{2} + \lambda_S  \frac{s^4}{4}
\\
&= \frac{1}{4} \lambda_H h^{4} +  \frac{1}{4} \lambda_{HS} h^{2} s^{2} + \frac{1}{4} \lambda_{S} s^{4}.
\end{split}
\label{eq:V}
\eeq
The couplings $ \lambda_{H}$, $\lambda_{HS} $, $ \lambda_S$ depend on the RG scale $\bar\mu$ according to the $\beta$-functions  $\beta_X = dX/dt$ with $t = \ln(\bar\mu^2/\bar\mu_0^2)/{(4\pi)^2}$. As we formulate the effective theory, we leave the $\beta$-functions generic.

The possible phases of dynamical symmetry breaking depend on which field acquires a VEV:
\begin{enumerate}
\item[$s$)] $s\neq 0$ and $h=0$ arises when the critical boundary
\beq 
\label{eq:betacritS}
\lambda_S =0
\eeq 
is crossed, while $\lambda_{HS}>0$ gives a tree-level positive squared mass to the Higgs boson and consequently the two scalars do not mix. Dynamical symmetry breaking happens if $\beta_{\lambda_S}>0$.

\item[$h$)] $h\neq 0$ and $s=0$ arises when $\lambda_H =0$ and $\lambda_{HS}>0$.
Similarly to the previous case, the two scalars do not mix as only one of them acquires a non-vanishing VEV. This is the scenario originally considered by Coleman and Weinberg, now excluded by the Higgs mass measurement that implies $\lambda_H \approx 0.13$.

\item[$sh$)] $s,h\neq 0$ appears when the critical boundary,
 \beq \label{eq:shcond}
2\sqrt{\lambda_H \lambda_S}+\lambda_{HS} =0,
\eeq 
is crossed, while $\lambda_{HS}<0$ and $\lambda_{H,S}\ge 0$.
The flat direction is  given by $s/h = (\lambda_H/\lambda_S)^{1/4}$.
Dynamical symmetry breaking happens if 
\beq 
\label{eq:betacrit}
\beta_{\rm crit}=\lambda_S \beta_{\lambda_H} + \lambda_H \beta_{\lambda_S} -\lambda_{HS} \beta_{\lambda_{HS}}/2 >0
\eeq 
along the critical boundary. 
In this phase the two mass eigenstates are superpositions of the original scalar fields.
\end{enumerate}

The phases $h)$ and $s)$ are not smoothly connected and 
the potential has two disjoint local minima with $h\neq 0$ and with $s\neq 0$, 
corresponding to a first-order phase transition with no extra light scalars~\cite{Kannike:2021iyh}.

On the other hand, the phases $s)$ and $sh)$ are smoothly connected, so the flat direction along the field $s$ can be deformed to yield a minimum in the $sh)$ phase. As the squared Higgs boson mass changes sign across the phases $s$) and $sh$), the Higgs boson is necessarily light near the multi-phase criticality point at their intersection. Furthermore, as the Higgs boson does not acquire a VEV in the phase $s)$, the scalar mixing is also naturally suppressed in proximity of the critical boundary. The two conditions in Eqs.~\eqref{eq:betacritS} and~\eqref{eq:shcond} intersect at 
\beq\label{eq:multiphase} \lambda_S(\bar\mu)=\lambda_{HS}(\bar\mu)=0, \eeq 
that trivially implies a massless Higgs boson.

We summarize the usual Gildener-Weinberg computation for the phase $sh)$ 
as it provides an example of how dynamical symmetry breaking can be approximated
using the RG-improved tree-level potential alone. For the sake of generality, we leave all the model-dependent one-loop contribution to the $\beta$-functions implicit.

The masses and mixings of scalars in this scenario can be obtained from the one-loop potential 
\begin{equation}
  V = V^{(0)} + V^{(1)},
\label{eq:V:eff}
\end{equation}
with the tree-level part $V^{(0)}$ given in Eq.~\eqref{eq:V} and having omitted terms involving other possible heavier scalar fields. The one-loop contribution, $V^{(1)}$, is given by
\beq \label{eq:V1}
V^{(1)}|_{\overline{\rm MS}} = \frac{1}{4(4\pi)^2} \Tr \bigg[M_S^4 \left( \ln \frac{M_S^2}{\bar\mu^2} -\frac32 \right)+\qquad\eeq
$$\qquad\qquad- 2 M_F^4 \left(\ln\frac{M_F^2}{\bar\mu^2} -\frac32 \right)+ 3 M_V^4 \left( \ln\frac{M_V^2}{\bar\mu^2}-\frac56 \right)\bigg],$$
where $\bar\mu$ indicates the RG scale introduced by dimensional regularization in the $\overline{\rm MS}$ scheme. The parameters in the tree-level part of the effective potential run as dictated by the Callan-Symanzik equation.
Their dependence on $\bar\mu$ thus cancels that of the one-loop contribution, making the effective potential independent of the arbitrary renormalization scale
up to higher-loop orders and wave-function renormalization. The symbols $M_{S,F,V}$ denote  the usual field-dependent masses of generic scalars, fermions and vectors, respectively.
For example, $M_V^2 = g_h^2 h^2 + g_s^2 s^2$ is the mass of the U(1) gauge boson in a model where $h$ and $s$ have corresponding gauge charges $g_h$ and $g_s$.

Along the tree-level flat direction, the potential can be approximated by expanding the tree-level term at the first order in the $\beta$-functions: 
 \be
 \lambda_{{\rm eff}, i}(s') = \lambda_{i}(s_{0}) + \Delta \lambda_{i}(s_{0}) + \beta_{\lambda_{i}}(s_{0}) \frac{1}{(4 \pi)^{2}}  \ln \frac{s^{\prime 2}}{s_0^2},
 \ee
 where $s_0$ is a typical scale (\eg, the flat direction scale) and $s^{\prime2} = s^2 + h^2$ is the distance in field space.
This approximation is appropriate along the flat direction, rather than in all field space. Because our tree-level flat direction is along the $s$-axis, we choose $s' = s$. The finite corrections $\Delta \lambda_{i}(s_{0})$ arise from a Taylor expansion of the one-loop term $V^{(1)}$ in powers of $h^{2}$. We stress that taking them into account -- whereby going beyond the usual Gildener-Weinberg approximation -- is crucial in order to obtain the correct values for the Higgs VEV and mass in the minimum.

Due to these corrections, the tree-level condition in Eq.~\eqref{eq:multiphase} does not hold any more. For convenience, we define the scales where $\lambda_{S}(s)$ and $\lambda_{HS}(s)$ cross zero by $s_{S}$ and $s_{HS}$, respectively. They can be obtained from 
\begin{align}
  0 &= \lambda_{S} + \Delta \lambda_{S} + \beta_{\lambda_{S}} \frac{1}{(4 \pi)^{2}}  \ln \frac{s_{S}^{2}}{s_0^2},
  \\
  0 &= \lambda_{HS} + \Delta \lambda_{HS} + \beta_{\lambda_{HS}} \frac{1}{(4 \pi)^{2}}  \ln \frac{s_{HS}^{2}}{s_0^2},
\end{align}
where all the quantities are given at the scale $s_{0}$. The precise order-one value of the quantity 
\begin{equation}
\label{eq:R}
  R = e^{-1/2} s_S^2/s_{HS}^2
\end{equation}
is model-dependent and encodes the deviation from the tree-level Gildener-Weinberg approximation which predicts $\ln R=-1/2$.

Eq.~\eqref{eq:V1} has been computed in Ref.~\cite{Kannike:2021iyh}  using simplifications appropriate around the multi-phase critical point, which result
in simple analytic expressions for the mixing, VEVs, and masses of the light particles. In more detail, assuming that the $\beta$-functions of $\lambda_S$ and $\lambda_{HS}$ are comparable and much smaller than $\lambda_H$,
the potential has a minimum at non-vanishing $s$ and $h$,
\beq s \approx e^{-1/4} s_S ,\qquad
h \approx 
\frac{ e^{-1/4} s_S}{4\pi} \sqrt{\frac{-\beta_{\lambda_{HS}}\ln R}{2\lambda_H}},
\label{eq:vevs}
\eeq
provided that $-\beta_{\lambda_{HS}}\ln R>0,$ otherwise only $s$ acquires a VEV.  The resulting mass eigenvalues are both loop-suppressed,
\beq 
\label{eq:masses}
m_s^2 \approx \frac{2s^2 \beta_{\lambda_S}}{(4\pi)^2},\qquad
m_h^2 \approx \frac{-s^2 \beta_{\lambda_{HS}}\ln R}{(4\pi)^2} = 2\lambda_H h^2 ,
\eeq
and the mixing angle is also loop-suppressed, barring degenerate scalar masses:
\beq
\label{eq:mixing}
\theta \approx \sqrt{-\frac{\beta_{\lambda_{HS}}^3 \ln R}{2\lambda_H}} 
\frac{1+\ln R}{4\pi(2\beta_{\lambda_S} + \beta_{\lambda_{HS}} \ln R)}.
\eeq
Possible mutual dependencies between the parameters $m_h, m_s,\theta$ can only be specified within the context of a complete model for the proposed mechanism. 

To summarize, the low energy effective model discussed above can be embedded in the SM by extending the particle content with one extra light scalar, the dilaton. The dilaton is weakly mixed with the Higgs boson and does not have other interactions with the remaining SM degrees of freedom. The parameters which can be measured at colliders are the masses of the two scalars and their mixing angle, given by Eq.~(\ref{eq:masses}) and Eq.~(\ref{eq:mixing}), respectively. The VEVs of the fields are determined by Eq.~\eqref{eq:vevs}. On the other hand, there are four parameters describing the multi-phase criticality scenario, $s$, $\lambda_{S}$, $\lambda_{HS}$ and $\ln R$, in addition to the already known Higgs boson parameters $h = 246.2$~GeV and $\lambda_H = 0.13$ (or, equivalently, $m_h = 125.1$~GeV). We observe that when connecting the low-energy observables to the parameters of the model, one quantity remains necessarily undetermined as the relevant equations all depend on the combination $\beta_{\lambda_{HS}} \ln R$. Therefore, the scale of new physics, $\Lambda\equiv s,$ can be model-independently determined from the low energy measurements only up to the logarithmic correction $\ln R.$ The Gildener-Weinberg approximation predicts $\ln R=-1/2$ and deviations from this value, although model dependent, are expected to be small for perturbative values of couplings. For example, in the three-scalar model considered in  Ref.~\cite{Kannike:2021iyh}, a precise computation gives $\ln R=-3/8$. In the following we shall use $\ln R=-1/2$ as the reference value, but model dependent corrections of order $\mathcal{O}(1)$ are possible.

\section{Collider Phenomenology}
\label{sec:coll}

In this Section we present in detail the collider analysis of the present scenario in the context of both hadron and lepton colliders. For the numerical computation, the model was implemented in Feynrules~\cite{Christensen:2008py, Degrande:2011ua, Alloul:2013bka}. The events were generated by using Madgraph5~\cite{Alwall:2011uj, Alwall:2014hca} and the subsequent 
showering was done using Pythia8~\cite{Sjostrand:2014zea}. 
The detector simulation was performed with Delphes3~\cite{deFavereau:2013fsa,Selvaggi:2014mya}. The jets are reconstructed with anti-kt algorithm~\cite{Cacciari:2008gp} by Fastjet~\cite{Cacciari:2011ma}.

\subsection*{LHC Constraints}
\label{sec:bp_choice}

\begin{figure}[h!]
\includegraphics[width=7.0cm,height=6cm]{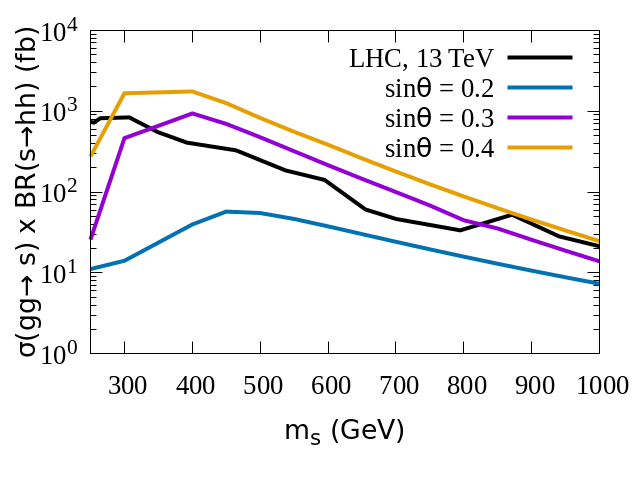}
\caption{The impact of the LHC direct searches on the dilaton parameter space for 
different choices of the mixing angle $\sin\theta$= 0.2, 0.3 and 0.4. The experimental exclusion line is obtained from 13 TeV run of the LHC~\cite{ATLAS:2019qdc}.} 
\label{fig:constraint}
\end{figure}

The mass of the dilaton and its mixing angle with the 125 GeV Higgs are constrained from the LHC search in various final states~\cite{ATLAS:2019qdc,CMS:2019bnu}. In the present scenario, the small mixing angle 
suppresses the production cross-section of the new scalar, hence a light scalar is still allowed if the mixing angle is small enough. We observed that the most stringent constraint comes from the combined $95\%$ 
confidence level limit targeting the production cross-section times branching ratio (BR) for the process $\sigma(gg\to s)\times {\rm ~BR}(s\to hh)$. The combined final states include $WWWW$, $WW\gamma\gamma$, $b\bar b\gamma\gamma$, $b\bar b b\bar b$ and $b\bar b\tau\bar\tau$~\cite{ATLAS:2019qdc}. In Fig.~\ref{fig:constraint} we show the impact of this result on our parameter space. 

Clearly, $m_s \lesssim 1000$ GeV is ruled out for $\sin\theta= 0.4$ except for a small region of parameter space where $m_s$ is close to the kinematic threshold for the decay $s\to hh$. For $\sin\theta$= 0.3, $m_s \lesssim 820$ GeV is ruled out except the region $250 < m_s \lesssim 340$ GeV since here BR($s\to hh$) is not large enough to produce enough yield for the signal events. In these regions $s\to VV$ where $V\equiv W,Z$ are the most dominant decay modes. Smaller mixing angles such as $\sin\theta$= 0.2 cannot be probed with the existing data.

\subsection{High-Luminosity LHC}
\label{sec:coll_had}

We then explore the following signal region in the context of $14$ TeV LHC. 
\begin{itemize}
\item {\bf SR1:} The dilaton is produced through gluon-gluon fusion and decays into two SM Higgs bosons. 
The two SM Higgs bosons decay into $b\bar b$ and $\gamma\gamma$, respectively, giving rise to a $2{~\rm b-jets} ~+~ 2\gamma ~+~ \slashed{E}_T$ final state.
\end{itemize}

\subsubsection*{High-Luminosity LHC Results for SR1}
\label{sec:coll_res_had}

The benchmark points we choose for this analysis are BP1 ($m_s = 600$ GeV, ${\rm sin}\theta = 0.2$) and BP2 ($m_s = 800$ GeV, ${\rm sin}\theta = 0.2$). Since the higher mixing angle values are mostly ruled out, we 
focus on probing smaller mixing angle. The dominant background channels we studied in this context 
are $t\bar t\gamma\gamma$, $b\bar b\gamma\gamma$, $t\bar t h~(h\to \gamma\gamma)$, 
$b\bar b h~(h\to \gamma\gamma)$, $\gamma\gamma~+~{\rm jets}$ and $Zh (h\to\gamma\gamma)$. 
We implement the following cuts to reduce the number of background events in order to 
increase our signal sensitivity.
\begin{itemize}
\item {C11:} The final state must consist of two $b$-jets with $p_T > 25$ GeV and two photons with $p_T > 20$ GeV. We veto all events containing charged leptons with $p_T > 20$ GeV.
\item {C12:} Unlike some of the background channels, the signal events do not have any direct source of missing energy. Hence, we restrict the transverse missing energy to $\slashed{E}_T < 80$ GeV. 
\item {C13:} The two $b$-jets are obtained from the SM Higgs decay, thus we restrict their invariant mass to $90 < m^{\rm inv}_{b\bar b} < 135$ GeV. 
\item {C14:} The effective mass of the final state has to satisfy the requirement 
$m_{\rm eff} > 500$ GeV. Here, 
$m_{\rm eff} = \sum p_T^{\ell_i} + \sum p_T^{j_i} + \sum p_T^{\gamma_i} + \slashed{E}_T$. 
\item {C15:} The two photons also arise from the decay of the SM Higgs and for the corresponding invariant mass we thus require $120 < m^{\rm inv}_{\gamma\gamma} < 130$ GeV. 
\item {C16:} The $b$-jet pair and the photon pair in the signal events are expected to be well separated. We restrict their angular separation to  $\Delta\phi_{b\bar b, \gamma\gamma} > 2.0$. 
\item {C17:} Finally, we impose the following $p_T$ criteria on the photons in the final state, 
$\frac{p_T^{\gamma_1}}{m^{\rm inv}_{\gamma\gamma}} > 0.33$ and 
$\frac{p_T^{\gamma_2}}{m^{\rm inv}_{\gamma\gamma}} > 0.25$. Here, $\gamma_1$ and $\gamma_2$ indicate the hardest and second hardest photons in the final state respectively.
\end{itemize} 

Table~\ref{tab:sr1_res} presents the cut-flow table for the SR1 analysis.

\begin{table}[h]
\small
\begin{tabular}{||c||c c c c c c c||}
\hline
\multicolumn{1}{||c||}{Channels} &
\multicolumn{7}{|c|}{Cross-section (fb)} \\
\cline{2-8}
 & C11  & C12 & C13 & C14 & C15 & C16 & C17    \\  
\hline\hline
Signal (BP1) & 0.023 & 0.022 & 0.016 & 0.013 & 0.013 & 0.012 & 0.010         \\
Signal (BP2) & 0.006 & 0.005 & 0.0041 & 0.0038 & 0.0037 & 0.0035 & 0.0033         \\
\cline{1-8}
$t\bar t\gamma\gamma$            & 7.016 & 4.868 & 1.081 & 0.302 & 0.010 & 0.005 & --  \\
$b\bar b\gamma\gamma$            & 33.68 & 33.36 & 7.922 & 0.154 & -- & --  & --   \\
$t\bar t h~(h\to \gamma\gamma)$  & 0.068 & 0.052 & 0.013 & 0.006 & 0.006 & 0.003 & 0.003   \\
$b\bar b h~(h\to \gamma\gamma)$  & 0.015 & 0.015 & 0.003 & 0.001 & 0.001 & -- & --   \\
$\gamma\gamma~+~{\rm jets}$      & 32.13 & 31.68 & 6.458 & 0.812 & 0.041 & -- & --    \\  
$Zh (h\to\gamma\gamma)$          & 0.010 & 0.010 & 0.002 & -- & -- & -- & --   \\ 
\hline\hline
\end{tabular}
\caption{Results of cut-based analysis for the sample benchmark points BP1 ($m_s=600$ GeV, $\sin\theta = 0.2$) 
and BP2 ($m_s=800$ GeV, $\sin\theta = 0.2$) corresponding to the signal region SR1 and the dominant background channels. 
The center-of-mass energy is taken to be $\sqrt{s}=14$ GeV. The signal events are produced via gluon-gluon fusion.}
\label{tab:sr1_res}
\end{table}

The luminosities required to obtain a $3\sigma$ statistical significance for BP1 and BP2 in SR1 are $\sim 1200~{\rm fb}^{-1}$ and $\sim 5200~{\rm fb}^{-1}$, respectively. Clearly, BP2 cannot be probed to its discovery significance at the LHC, but it is possible to obtain a $2\sigma$ indication at a luminosity of $\sim 2300~{\rm fb}^{-1}$.  For $m_s\gtrsim 1$ TeV, the parameter space 
cannot be probed with good sensitivity even for $\sin\theta = 0.3$.

The available parameter space just above $m_s = 250$ GeV cannot be probed at the 14 TeV LHC since the signal cross-section is suppressed by small branching ratio and smaller cut efficiencies. In the next Section, we show that this parameter space can be explored with better efficiency at a lepton collider.  

\subsection{Future Lepton Collider}
\label{sec:coll_lep}

In the context of a lepton collider, we explore the expected phenomenology with two different center-of-mass energies: 500 GeV and 1 TeV. At $\sqrt{s}=500$ GeV, the dilaton production cross-section is most dominant in $ZH$ and vector boson fusion (VBF) modes and the cross-sections are comparable~\cite{Abramowicz:2016zbo}. Above this center-of-mass energy, VBF dominates over the other production cross-sections. Hence in this work we concentrate only on the VBF production of the dilaton. We explore two different 
signal regions delineated by the possible decay patterns.
\begin{itemize}
\item {\bf SR2:} The dilaton is produced along with two neutrinos and subsequently decays into two $W$-bosons. We consider the two $W$-bosons decaying leptonically and hadronically giving rise to a 
signal: $1\ell ~+~2{~\rm jets} ~+~\slashed{E}$. This channel is explored at $\sqrt{s}=500$ GeV.
\item {\bf SR3:} After being produced via VBF, the dilaton decays into two SM Higgs states. These then decay into a $b\bar b$ pair and a $WW^*$ pair, resulting in a $2b{~\rm jets} ~+~ 2{~\rm fat~jets} ~+~ \slashed{E}$ final state. The two fat jets originate from the two gauge boson decays. This channel is investigated at $\sqrt{s}=1$ TeV. We remark that leptonic decay(s) of the gauge boson(s) in the present signal region lead to a signal cross-section too small to be observed even at feasible luminosities. 
\end{itemize}

\subsubsection*{Lepton Collider Results for SR2}
\label{sec:coll_res_lep}

This analysis considers a center-of-mass energy of $\sqrt{s}=500$ GeV. 
The benchmark point (BP3) we choose to study this signal region is specified by $m_s = 250$ GeV and $\sin\theta = 0.3$. The $s$ scalar decays dominantly into $WW$ with a corresponding branching ratio of $\sim 70\%$ for
$\sin\theta = 0.25$. Smaller mixing angles result in too 
small cross-sections. For larger values as $\sin\theta=0.3,~0.4$, 
BR($s\to WW$) decreases and again the signal cross-section drops. Hence 
$\sin\theta \sim 0.3$ is the optimum value for which the best sensitivity is obtained considering $m_s = 250$ GeV. The dominant background channels for the $1\ell ~+~2{~\rm jets} ~+~\slashed{E}$ final state are $VV$, $VVV$ 
and $Zh$, where, $V\equiv W, Z$, which we suppress through the following kinematical cuts.
\begin{itemize}
\item{C21:} The signal region must have only one lepton with $p_T > 15$ GeV and 
$|\eta| < 2.3$. The two jets must have $p_T > 30$ GeV and $|\eta| < 4.7$. The $p_T$ of the 
final state lepton and jets are further restricted to $p_T^{\ell} < 120$ GeV, $p_T^{j_1} < 100$ GeV and $p_T^{j_2} < 60$ GeV.
\item {C22:} The invariant mass of the two jets should peak around the $W$ boson mass. Hence $70 < M^{\rm inv}_{jj} < 90$. 
\item{C23:} The separation between two jets in the final state must be small since they are both originating from one single $W$ boson. Hence, $1.0 < \Delta R_{jj} < 2.0$. 
\item{C24:} The missing energy of the final state should be large: $280 < \slashed{E} < 400$ GeV. 
\item{C25:} The energy of the $W$-candidate decaying into two jets must obey $100 < E_W < 160$ GeV.   
\item{C26:} The angular separation between the charged lepton and missing energy vector is in the range $\Delta\phi_{\ell,\slashed{E}} < 2.0$.
\end{itemize} 

Table~\ref{tab:sr2_res} represents the cut-flow table for the SR2 analysis.

\begin{table}[h!]
\small
\begin{tabular}{||c||c c c c c c||}
\hline
\multicolumn{1}{||c||}{Channels} &
\multicolumn{6}{|c|}{Cross-section (fb)} \\
\cline{2-7}
 & C21  & C22 & C23 & C24 & C25 & C26   \\  
\hline\hline
Signal (BP3) & 0.077 & 0.055 & 0.048 & 0.034 & 0.031 & 0.024        \\
\cline{1-7}
$WW$          & 287.9 & 207.2 & 126.7 & 0.096 & 0.048 & 0.024      \\
$WWZ$         & 1.73 & 0.643 & 0.414 & 0.149 & 0.112 & 0.080        \\
$ZZZ$         & 0.011 & 0.004 & 0.002 & 0.001 & 0.001 & -- \\
$ZZ$          & 3.89 & 2.076 & 0.742 & 0.004 & --  & -- \\
$Zh$          & 0.173 & 0.058 & 0.049 & 0.014 & 0.009 & 0.001 \\      
\hline\hline
\end{tabular}
\caption{Results of cut-based analysis for the sample benchmark point BP3 ($m_s=250$ GeV, $\sin\theta = 0.3$) corresponding to the signal region SR2 and the dominant background channels. The centre-of-mass energy is taken to be $\sqrt{s}=500$ GeV. The signal events are produced via VBF.}
\label{tab:sr2_res}
\end{table}

The  luminosity required to obtain a $3\sigma$ statistical significance for BP2 in SR2 is $\sim 2050~{\rm fb}^{-1}$.

\subsubsection*{Lepton Collider Results for SR3}

This analysis is performed with a centre-of-mass energy of $\sqrt{s}=1$ TeV.
The benchmark points we choose, BP4 and BP5, have $m_s=500$ GeV, $\sin\theta=0.2$ and 
$m_s=600$ GeV, $\sin\theta=0.2$,  respectively. In this case $s\to hh$ is the most dominant decay mode, with a branching ratio of $\sim 81\%$
and $\sim 94\%$ respectively for BP4 and BP5. The dominant background 
channels for the $2{~\rm b-jets} ~+~ 2{~\rm fat~jets} ~+~ \slashed{E}$ final state are $t\bar t + {\rm jets}$, $VVV$ and $Zh$. The following kinematical cuts are imposed in order to reduce the background contribution.

\begin{itemize}
\item {C31:} The signal region must have two $b$-jets with $p_T > 30$ GeV and $|\eta| < 4.7$.  
The signal region also must consist of exactly two fat-jets (constructed with R parameter=1.0) with $p_T > 30$ GeV and $|\eta| < 4.7$. We impose a veto on leptons in the final state with $p_T^{\ell} > 15$ GeV and $|\eta|^{\ell} < 2.3$.  
\item {C32:} The missing energy is $400 < \slashed{E} < 700$. 
\item {C33:} The invariant mass of the fat-jet pair are restricted to  
$90 < m^{\rm inv}_{JJ} < 130$ since they originate from the 125 GeV Higgs in the signal events.
\item {C34:} The $b$-jet pair must be close to each other, $\Delta R_{b\bar b} < 2.5$, and so should the fat-jet pair, $\Delta R_{JJ} < 2.5$.   
\item {C35:} The invariant mass of the $b$-jet pair must peak at the Higgs mass, hence we restrict the parameter to $90 < m^{\rm inv}_{b\bar b} < 130$. 
\end{itemize}

Table~\ref{tab:sr3_res} represents the cut-flow table for the SR3 analysis.

\begin{table}[h!]
\small
\begin{tabular}{||c||c c c c c||}
\hline
\multicolumn{1}{||c||}{Channels} &
\multicolumn{5}{|c|}{Cross-section (fb)} \\
\cline{2-6}
 & C31  & C32 & C33 & C34 & C35     \\  
\hline\hline
Signal (BP4)  & 0.020 & 0.019 & 0.011 & 0.009 & 0.008        \\
Signal (BP5)  & 0.022 & 0.019 & 0.011 & 0.008 & 0.007     \\                          
\cline{1-6}
$t\bar t$     & 4.73 & 0.171 & 0.036 & 0.007 & 0.001              \\
$WWZ$         & 0.395 & 0.046 & 0.005 & 0.001 & 0.0003               \\
$ZZZ$         & 0.017 & 0.002 & 0.001 & 0.0005 & 0.0001          \\
$Zh$          & 0.006 & $< 10^{-4}$ & -- & -- & --          \\
\hline\hline
\end{tabular}
\caption{Results of cut-based analysis for the sample benchmark point BP4 ($m_s=500$ GeV, $\sin\theta = 0.2$) corresponding to the signal region SR3 and the dominant background channels. The centre-of-mass energy is taken to be $\sqrt{s}=1$ TeV. The signal events are produced via VBF.}
\label{tab:sr3_res}
\end{table}

Evidently, the cuts are quite efficient in suppressing the background events and, even though the signal cross-section is not very large, it is possible to obtain a good sensitivity. The luminosities required to reach the $3\sigma$ statistical significance for BP4 and BP5 in SR3 are $\sim 1330~{\rm fb}^{-1}$ and $\sim 1550~{\rm fb}^{-1}$, respectively.  

\subsubsection*{Impact of Polarization}

One advantage of a lepton-lepton collider is the possibility of polarising the incoming beams to enhance the signal cross-section. For example, as per CLIC design, the electron beams can be polarized up to $\pm 80\%$. The positron beam can also be polarized at a lower level. In our case, the electron-positron polarization of $-80\%~:~+30\%$  enhances the $e^+ e^-\to H\nu\bar\nu$
cross-section by a factor of $\sim 2.34$~\cite{Abramowicz:2016zbo}, owing to the two vertices in the VBF diagram involving an electron, a $W$-boson and a neutrino. However, the cross-sections of some of the background channels are also similarly enhanced. We checked that the cross-sections of the background channels $WW$ and $WWZ$ increase by a factor $\sim 2.3$. Among the other channels, the cross-sections of $t\bar t$, $ZZ$ and $ZZZ$ increase by 
a factor of $\sim 1.8$, while the cross-section of $Zh$ channel increases by a factor $\sim 1.5$. In table~\ref{tab:res_epem_pol} we present an estimate of how much the signal sensitivity may increase by exploiting the beam polarization. 

\begin{table}[h!]
\small
\begin{tabular}{||c||c|c||c|c||}
\hline
\multicolumn{1}{||c||}{Benchmark} &
\multicolumn{4}{|c|}{Required Luminosity fb$^{-1}$} \\
\cline{2-5}
\multicolumn{1}{||c||}{} &
\multicolumn{2}{|c||}{SR2} &
\multicolumn{2}{|c|}{SR3}  \\
\cline{2-5}
Points & No Pol. & Pol. & No Pol. & Pol.     \\  
\hline\hline
BP3  &$\sim 2050$ &$\sim 850$ & -- & --        \\
\hline
BP4  & -- & -- & $\sim 1330$ &  $\sim 550$       \\  
BP5  & -- & -- & $\sim 1550$ &  $\sim 640$       \\ 
\hline\hline
\end{tabular}
\caption{Luminosity in fb$^{-1}$ required for obtaining a $3\sigma$ statistical significance for BP3, BP4 and BP5 in signal regions SR2 and SR3, respectively with non-polarized (No Pol.) and polarized (Pol.) lepton beams. The electron and positron polarizations are taken to be $-80\%~:~+30\%$. }
\label{tab:res_epem_pol}
\end{table}

As we can see, the polarization effect improves the sensitivity quite significantly. The results suggest that the required luminosity decreases by a factor of $\sim 2.4$ when using polarized beams. 

\subsection{Sensitivity of Future Experiments}

\begin{figure}[h!]
\includegraphics[width=7.0cm,height=6cm]{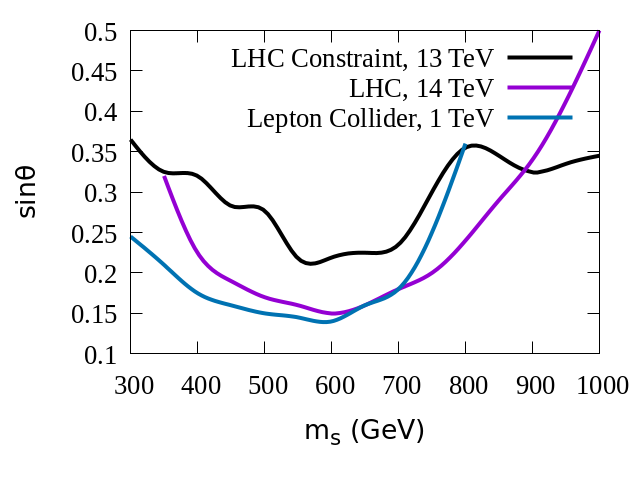}
\caption{The contours correspond to the exclusion bounds obtained at a 3$\sigma$ statistical significance in the context of the 14 TeV LHC run and 1 TeV future lepton collider, respectively for the signal regions SR1 (violet line) and SR3 (blue line). Both cases assume 3000 ${\rm~fb}^{-1}$ of luminosity. The black line represents the current 13 TeV LHC exclusion limit derived from the 36.1 ${\rm~fb}^{-1}$ luminosity results.} 
\label{fig:disc_lim}
\end{figure}

In this Section we show which part of the dilaton parameter space can be probed at the 14 TeV LHC and at a 1 TeV lepton collider. To this purpose, we employ the best setups identified in our previous analysis, corresponding to SR1, for the case of LHC, and SR3 for the lepton collider. In both cases, we use the maximum possible luminosity of 3000 ${\rm~fb}^{-1}$ and bound the parameters of the model at a 3$\sigma$ statistical significance. For the lepton collider case,  we have considered the polarized beam option mentioned in the previous Section as it provides a better sensitivity. The results obtained are presented in 
Fig.~\ref{fig:disc_lim}, which shows the exclusion region in $m_s$ - $\sin\theta$ plane for 3$\sigma$ significance,  along with the current bound due to the 13 TeV LHC data~\cite{ATLAS:2019qdc}.  

Clearly, for the kinematical cuts used in this study, the region $400\lesssim m_s\lesssim 800$ GeV can be probed with the most sensitivity at the 14 TeV LHC. The lepton collider presents a smaller reach in mass because of the centre-of-mass energy, but it allows to probe smaller mixing angles,
$\sin\theta \sim 0.15$, for dilaton masses below $m_s\simeq 600$ GeV. The values of mixing angle that can be probed for $m_s\gtrsim 900$ GeV and $m_s\gtrsim 800$ GeV in SR1 and SR3, respectively, are already excluded by 
the current LHC data.

\subsection{Implications for the scale of Symmetry Breaking}
\label{scale}

The results obtained for the power of current and future collider experiments  can be used to infer a lower bound on the scale where the dynamics of symmetry breaking takes place. 

\begin{figure}[h]
  \centering
  \includegraphics[width=0.4\textwidth]{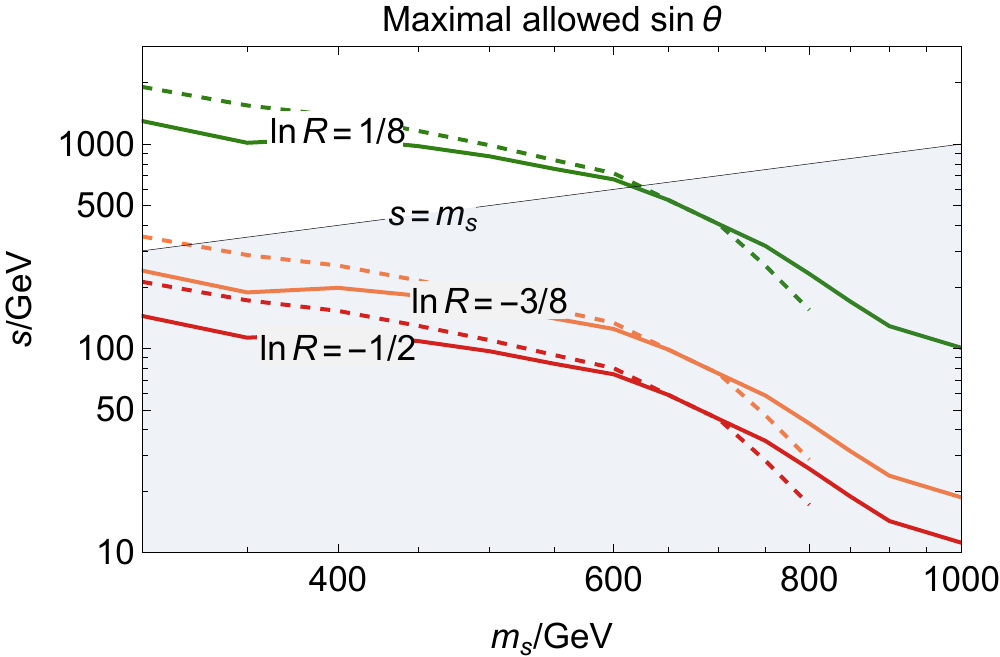}
  \caption{\label{fig:s}\em 
  The inferred scale of new physics probed at current and future collider experiments as a function of the dilaton mass. The colors indicate the solutions obtained for the indicated value of $\ln R$ using the maximal value allowed by the $3\sigma$ bound achievable at the LHC (solid lines), or the  projection for the reach of a future lepton collider (dashed lines). The shaded area indicates a region of the parameter space where perturbative unitarity cannot be guaranteed.}
  \end{figure}

\begin{figure}[h]
  \centering
  \includegraphics[width=0.4\textwidth]{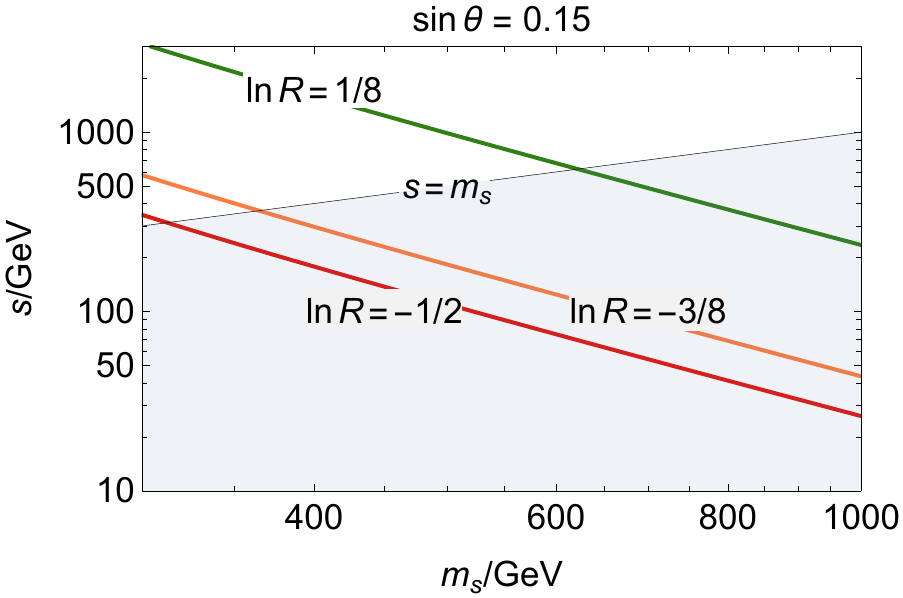}
  \caption{\label{fig:s2}\em 
  The inferred scale of new physics for a fixed value of the mixing angle,  $\sin\theta=0.15$, as a function of the dilaton mass and the model dependent correction, $\ln R$, to the Gildener-Weinberg solution. The shaded region denotes a part of the parameter space where perturbative unitarity of quartic couplings cannot be guaranteed. }
\end{figure}

In Fig.~\ref{fig:s} we show the lower bound on the expected new physics scale, $\Lambda\sim s$, obtained by inverting Eqs.~\eqref{eq:masses} and~\eqref{eq:mixing} for the range of dilaton masses and corresponding maximal mixing angle allowed shown in Fig.~\ref{fig:disc_lim}. As the collider observables can only determine the parameters of the theory up to the (model dependent) logarithmic correction $\ln R$, we show with different colors the solutions obtained by varying the parameter around the Gildener-Weinberg value $\ln R = - 1/2$. In each case, the dashed lines use the projections obtained for the maximal mixing angle allowed by future lepton colliders (blue line in Fig.~\ref{fig:disc_lim}), whereas the solid lines use the current and projected LHC data (the lowest between the purple and black lines of Fig.~\ref{fig:disc_lim}). The shaded area shows the region of parameter space where $m_S > s$, indicating through Eq.~\eqref{eq:masses} the potential loss of perturbative unitarity of the involved quartic coupling. This exercise is repeated in Fig.~\ref{fig:s2} for a fixed value of the mixing angle $\sin\theta=0.15$, with the borderline sensitivity expected for future colliders.

As we can see, the sensitivity of the considered collider experiment to the mixing angle is too low to result in a lower bound on the scale of new physics that significantly differs from the scale of the light scalar sector. In particular, the smallest value of mixing angle testable, $\sin\theta=0.15$, results only in a constraint that is, at most, one order of magnitude larger than the dilaton mass scale. We therefore conclude that collider experiments can constrain the scenario only marginally, including the case of a 1 TeV lepton collider that has the largest sensitivity to the parameter.

\section{Conclusions}
\label{concl}

In this work we studied the sensitivity of hadron and lepton colliders to the scale of new physics in models where
the lightness of the SM Higgs boson is explained by the multi-phase criticality phenomenon. In these models
the dynamical symmetry breaking scale can be inferred from the mixing between the two light scalar degrees of freedom --
the Higgs boson and the dilaton. Our results on the new physics scale are derived from studies of the latter at present and future colliders.

The results are collected in Fig.~\ref{fig:disc_lim}, which shows that present and future collider experiments are only sensitive to relatively large values of the mixing angle between the Higgs boson and the dilaton. As shown in Figs.~\ref{fig:s} and~\ref{fig:s2}, this causes a loss of sensitivity to the scale of new physics where symmetry breaking takes place, leading to a lower bound that does not significantly constrain the scenario. 
We therefore conclude that a high-luminosity lepton collider such as CLIC \cite{CLICdp:2018cto} or high-luminosity hadron collider such as 100 TeV FCC \cite{FCC:2018byv,FCC:2018evy,FCC:2018vvp} is needed to fully test our scenario.

\medskip

\paragraph*{Acknowledgements}
We thank A. Strumia for discussions. This work was supported by European Regional Development Fund through the CoE program grant TK133, the Mobilitas Pluss grants MOBTT5, MOBTT86, and the Estonian Research Council grants PRG434, PRG803 and PRG356.

\bibliographystyle{hunsrt}
\bibliography{artCW}

\begin{thebibliography}{10}

\bibitem{Wess:1974tw}
J.~Wess and B.~Zumino.
\newblock {Supergauge Transformations in Four-Dimensions}.
\newblock {\em Nucl. Phys. B}, 70:39--50, 1974.

\bibitem{Fayet:1977vd}
Pierre Fayet.
\newblock {Mixing Between Gravitational and Weak Interactions Through the
  Massive Gravitino}.
\newblock {\em Phys. Lett. B}, 70:461, 1977.

\bibitem{Chatrchyan:2012ufa}
Serguei Chatrchyan et~al.
\newblock {Observation of a New Boson at a Mass of 125 GeV with the CMS
  Experiment at the LHC}.
\newblock {\em Phys. Lett. B}, 716:30--61, 2012, 1207.7235.

\bibitem{Aad:2012tfa}
Georges Aad et~al.
\newblock {Observation of a new particle in the search for the Standard Model
  Higgs boson with the ATLAS detector at the LHC}.
\newblock {\em Phys. Lett. B}, 716:1--29, 2012, 1207.7214.

\bibitem{Barbieri:1999tm}
Riccardo Barbieri and Alessandro Strumia.
\newblock {What is the limit on the Higgs mass?}
\newblock {\em Phys. Lett. B}, 462:144--149, 1999, hep-ph/9905281.

\bibitem{Barbieri:2000gf}
Riccardo Barbieri and Alessandro Strumia.
\newblock {The 'LEP paradox'}.
\newblock In {\em {4th Rencontres du Vietnam}: {Physics at Extreme Energies
  (Particle Physics and Astrophysics)}}, 7 2000, hep-ph/0007265.

\bibitem{ArkaniHamed:2001nc}
Nima Arkani-Hamed, Andrew~G. Cohen, and Howard Georgi.
\newblock {Electroweak symmetry breaking from dimensional deconstruction}.
\newblock {\em Phys. Lett. B}, 513:232--240, 2001, hep-ph/0105239.

\bibitem{ArkaniHamed:2002qy}
N.~Arkani-Hamed, A.~G. Cohen, E.~Katz, and A.~E. Nelson.
\newblock {The Littlest Higgs}.
\newblock {\em JHEP}, 07:034, 2002, hep-ph/0206021.

\bibitem{Cheng:2003ju}
Hsin-Chia Cheng and Ian Low.
\newblock {TeV symmetry and the little hierarchy problem}.
\newblock {\em JHEP}, 09:051, 2003, hep-ph/0308199.

\bibitem{Chacko:2005pe}
Z.~Chacko, Hock-Seng Goh, and Roni Harnik.
\newblock {The Twin Higgs: Natural electroweak breaking from mirror symmetry}.
\newblock {\em Phys. Rev. Lett.}, 96:231802, 2006, hep-ph/0506256.

\bibitem{Kannike:2021iyh}
Kristjan Kannike, Luca Marzola, Martti Raidal, and Alessandro Strumia.
\newblock {Light Higgs boson from multi-phase criticality in dynamical symmetry
  breaking}.
\newblock {\em Phys. Lett. B}, 816:136241, 2021, 2102.01084.

\bibitem{Coleman:1973jx}
Sidney~R. Coleman and Erick~J. Weinberg.
\newblock {Radiative Corrections as the Origin of Spontaneous Symmetry
  Breaking}.
\newblock {\em Phys. Rev. D}, 7:1888--1910, 1973.

\bibitem{Gildener:1976ih}
Eldad Gildener and Steven Weinberg.
\newblock {Symmetry Breaking and Scalar Bosons}.
\newblock {\em Phys. Rev. D}, 13:3333, 1976.

\bibitem{Degrassi:2012ry}
Giuseppe Degrassi, Stefano Di~Vita, Joan Elias-Miro, Jose~R. Espinosa, Gian~F.
  Giudice, Gino Isidori, and Alessandro Strumia.
\newblock {Higgs mass and vacuum stability in the Standard Model at NNLO}.
\newblock {\em JHEP}, 08:098, 2012, 1205.6497.

\bibitem{Buttazzo:2013uya}
Dario Buttazzo, Giuseppe Degrassi, Pier~Paolo Giardino, Gian~F. Giudice,
  Filippo Sala, Alberto Salvio, and Alessandro Strumia.
\newblock {Investigating the near-criticality of the Higgs boson}.
\newblock {\em JHEP}, 12:089, 2013, 1307.3536.

\bibitem{Froggatt:1995rt}
C.~D. Froggatt and Holger~Bech Nielsen.
\newblock {Standard model criticality prediction: Top mass 173 +- 5-GeV and
  Higgs mass 135 +- 9-GeV}.
\newblock {\em Phys. Lett. B}, 368:96--102, 1996, hep-ph/9511371.

\bibitem{Linde:1983gd}
Andrei~D. Linde.
\newblock {Chaotic Inflation}.
\newblock {\em Phys. Lett. B}, 129:177--181, 1983.

\bibitem{Bezrukov:2007ep}
Fedor~L. Bezrukov and Mikhail Shaposhnikov.
\newblock {The Standard Model Higgs boson as the inflaton}.
\newblock {\em Phys. Lett. B}, 659:703--706, 2008, 0710.3755.

\bibitem{Akrami:2018odb}
Y.~Akrami et~al.
\newblock {Planck 2018 results. X. Constraints on inflation}.
\newblock {\em Astron. Astrophys.}, 641:A10, 2020, 1807.06211.

\bibitem{Christensen:2008py}
Neil~D. Christensen and Claude Duhr.
\newblock {FeynRules - Feynman rules made easy}.
\newblock {\em Comput. Phys. Commun.}, 180:1614--1641, 2009, 0806.4194.

\bibitem{Degrande:2011ua}
Celine Degrande, Claude Duhr, Benjamin Fuks, David Grellscheid, Olivier
  Mattelaer, and Thomas Reiter.
\newblock {UFO - The Universal FeynRules Output}.
\newblock {\em Comput. Phys. Commun.}, 183:1201--1214, 2012, 1108.2040.

\bibitem{Alloul:2013bka}
Adam Alloul, Neil~D. Christensen, C\'eline Degrande, Claude Duhr, and Benjamin
  Fuks.
\newblock {FeynRules 2.0 - A complete toolbox for tree-level phenomenology}.
\newblock {\em Comput. Phys. Commun.}, 185:2250--2300, 2014, 1310.1921.

\bibitem{Alwall:2011uj}
Johan Alwall, Michel Herquet, Fabio Maltoni, Olivier Mattelaer, and Tim
  Stelzer.
\newblock {MadGraph 5 : Going Beyond}.
\newblock {\em JHEP}, 06:128, 2011, 1106.0522.

\bibitem{Alwall:2014hca}
J.~Alwall, R.~Frederix, S.~Frixione, V.~Hirschi, F.~Maltoni, O.~Mattelaer,
  H.~S. Shao, T.~Stelzer, P.~Torrielli, and M.~Zaro.
\newblock {The automated computation of tree-level and next-to-leading order
  differential cross sections, and their matching to parton shower
  simulations}.
\newblock {\em JHEP}, 07:079, 2014, 1405.0301.

\bibitem{Sjostrand:2014zea}
Torbj\"orn Sj\"ostrand, Stefan Ask, Jesper~R. Christiansen, Richard Corke,
  Nishita Desai, Philip Ilten, Stephen Mrenna, Stefan Prestel, Christine~O.
  Rasmussen, and Peter~Z. Skands.
\newblock {An introduction to PYTHIA 8.2}.
\newblock {\em Comput. Phys. Commun.}, 191:159--177, 2015, 1410.3012.

\bibitem{deFavereau:2013fsa}
J.~de~Favereau, C.~Delaere, P.~Demin, A.~Giammanco, V.~Lema\^\i{}tre,
  A.~Mertens, and M.~Selvaggi.
\newblock {DELPHES 3, A modular framework for fast simulation of a generic
  collider experiment}.
\newblock {\em JHEP}, 02:057, 2014, 1307.6346.

\bibitem{Selvaggi:2014mya}
Michele Selvaggi.
\newblock {DELPHES 3: A modular framework for fast-simulation of generic
  collider experiments}.
\newblock {\em J. Phys. Conf. Ser.}, 523:012033, 2014.

\bibitem{Cacciari:2008gp}
Matteo Cacciari, Gavin~P. Salam, and Gregory Soyez.
\newblock {The anti-$k_t$ jet clustering algorithm}.
\newblock {\em JHEP}, 04:063, 2008, 0802.1189.

\bibitem{Cacciari:2011ma}
Matteo Cacciari, Gavin~P. Salam, and Gregory Soyez.
\newblock {FastJet User Manual}.
\newblock {\em Eur. Phys. J. C}, 72:1896, 2012, 1111.6097.

\bibitem{ATLAS:2019qdc}
Georges Aad et~al.
\newblock {Combination of searches for Higgs boson pairs in $pp$ collisions at
  $\sqrt{s} = $13 TeV with the ATLAS detector}.
\newblock {\em Phys. Lett. B}, 800:135103, 2020, 1906.02025.

\bibitem{CMS:2019bnu}
Albert~M Sirunyan et~al.
\newblock {Search for a heavy Higgs boson decaying to a pair of W bosons in
  proton-proton collisions at $\sqrt{s} =$ 13 TeV}.
\newblock {\em JHEP}, 03:034, 2020, 1912.01594.

\bibitem{Abramowicz:2016zbo}
H.~Abramowicz et~al.
\newblock {Higgs physics at the CLIC electron\textendash{}positron linear
  collider}.
\newblock {\em Eur. Phys. J. C}, 77(7):475, 2017, 1608.07538.

\bibitem{CLICdp:2018cto}
T.~K. Charles et~al.
\newblock {The Compact Linear Collider (CLIC) - 2018 Summary Report}.
\newblock 2/2018, 12 2018, 1812.06018.

\bibitem{FCC:2018byv}
A.~Abada et~al.
\newblock {FCC Physics Opportunities}: {Future Circular Collider Conceptual
  Design Report Volume 1}.
\newblock {\em Eur. Phys. J. C}, 79(6):474, 2019.

\bibitem{FCC:2018evy}
A.~Abada et~al.
\newblock {FCC-ee: The Lepton Collider}: {Future Circular Collider Conceptual
  Design Report Volume 2}.
\newblock {\em Eur. Phys. J. ST}, 228(2):261--623, 2019.

\bibitem{FCC:2018vvp}
A.~Abada et~al.
\newblock {FCC-hh: The Hadron Collider}: {Future Circular Collider Conceptual
  Design Report Volume 3}.
\newblock {\em Eur. Phys. J. ST}, 228(4):755--1107, 2019.

\end{thebibliography}

\end{document}